\documentclass[apl,preprint,reprint,superscriptaddress]{revtex4-1}
\usepackage{graphicx}
\usepackage{dcolumn}
\usepackage{bm}
\usepackage{epstopdf}
\graphicspath{{./images/}}

\usepackage[version=3]{mhchem}
\usepackage{graphicx}
\usepackage{dcolumn}
\usepackage{bm}
\usepackage{epstopdf}
\graphicspath{{./images/}}

\begin{document}

\title{Amorphous silicon-doped titania films for on-chip photonics}

\author{Thomas Kornher}
\email{t.kornher@physik.uni-stuttgart.de}
\author{Kangwei Xia}

\author{Roman Kolesov}

\affiliation{3. Physikalisches Institut, Universit\"{a}t Stuttgart, 70569 Stuttgart, Germany}
\author{Bruno Villa}

\affiliation{Semiconductor Physics Group,Cavendish Laboratory, JJ Thomson Avenue, Cambridge,CB3 0HE,UK}
\author{Stefan Lasse}

\affiliation{3. Physikalisches Institut, Universit\"{a}t Stuttgart, 70569 Stuttgart, Germany}
\author{Cosmin S. Sandu}

\affiliation{3D-OXIDES, 130 Rue Gustave Eiffel, Saint Genis Pouilly, 01630, France}
\author{Estelle Wagner}

\affiliation{3D-OXIDES, 130 Rue Gustave Eiffel, Saint Genis Pouilly, 01630, France}
\author{Scott Harada}

\affiliation{3D-OXIDES, 130 Rue Gustave Eiffel, Saint Genis Pouilly, 01630, France}
\author{Giacomo Benvenuti}

\affiliation{3D-OXIDES, 130 Rue Gustave Eiffel, Saint Genis Pouilly, 01630, France}
\author{Hans-Werner Becker}

\affiliation{RUBION, Ruhr-Universit\"{a}t Bochum, 44780 Bochum, Germany}
\author{J\"{o}rg Wrachtrup}

\affiliation{3. Physikalisches Institut, Universit\"{a}t Stuttgart, 70569 Stuttgart, Germany}


\begin{abstract}
High quality optical thin film materials form a basis for 
on-chip photonic micro- and nano-devices, where several photonic elements form
an optical circuit. Their realization generally requires the thin 
film to have a higher refractive index than the substrate material. 
Here, we demonstrate a method of depositing amorphous
25\% Si doped TiO$_2$ films on various substrates, a way of shaping these films into 
photonic elements, such as optical waveguides and resonators, and finally, the performance 
of these elements. The quality of the film is estimated by measuring thin film cavity
Q-factors in excess of $10^{5}$ at a wavelength of 790\,nm, 
corresponding to low propagation losses of 5.1\,db/cm. 
The fabricated photonic structures were used to optically address 
chromium ions embedded in the substrate by evanescent coupling, therefore enabling it 
through film-substrate interaction.
Additional functionalization of the films by doping with optically active rare-earth ions
such as erbium is also demonstrated. 
Thus, Si:TiO$_2$ films allow for creation of high quality photonic elements, both passive and
active and also provide access to a broad range of substrates and emitters embedded therein. 

\end{abstract}

\maketitle

\begin{figure}[h]
\includegraphics[width=0.5\textwidth]{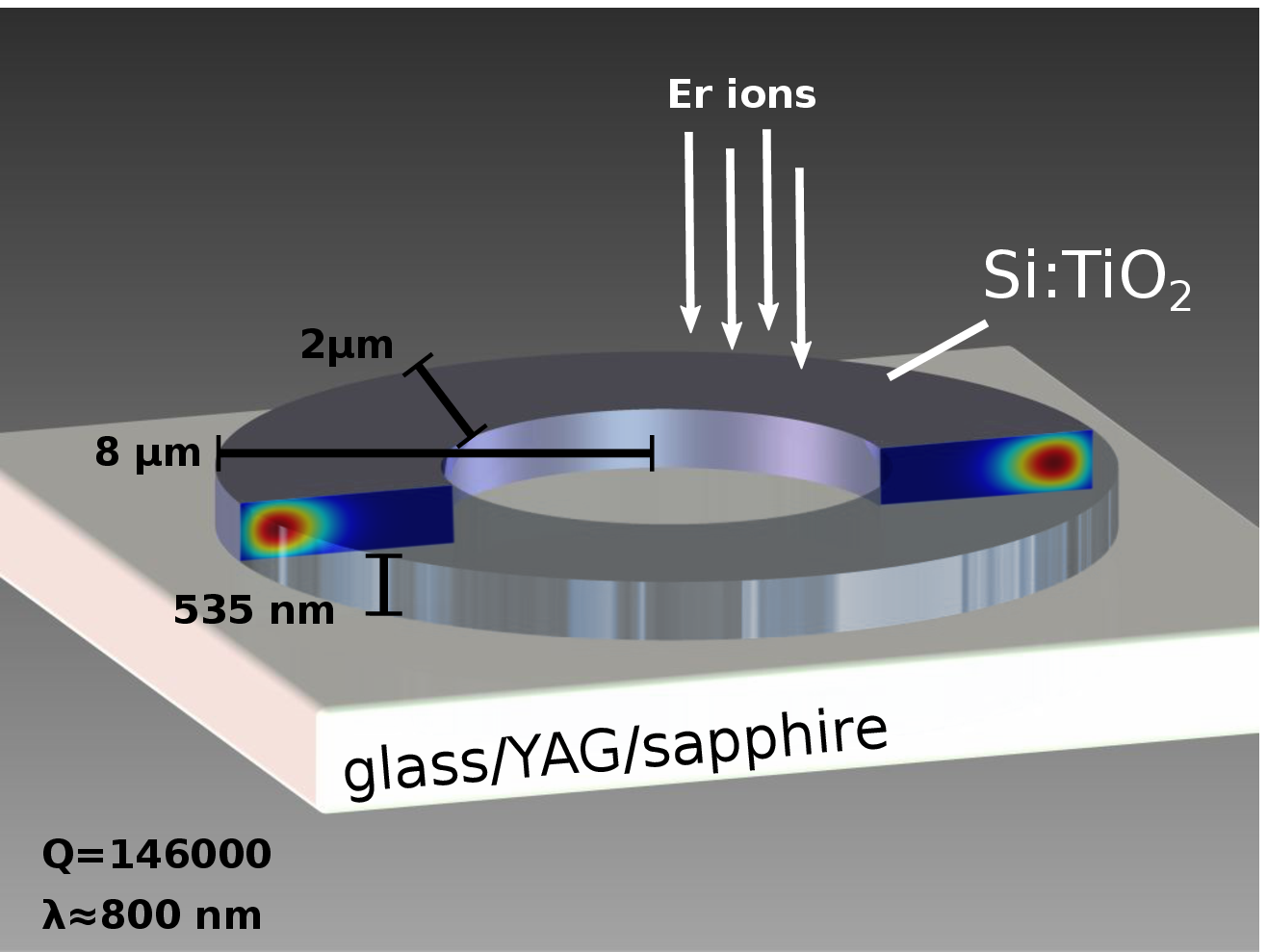}% Here is how to import EPS
\end{figure}

\textbf{Introduction}\\

Nanoscale fabrication of optical waveguide structures offers a wide range of 
opportunities spanning from on-chip photonic devices for optical networks to
sensor applications.
At the heart of such devices commonly lies a certain functionality that can be readily interfaced
with other components, and be included into sophisticated architectures as it is standard with
silicon-on-chip technology \cite{hulme2015fully}. Material systems hardly compatible with
silicon photonics,
such as YAG, YSO or sapphire crystals, can host a large variety of emitters with 
applications ranging from quantum communication 
\cite{kolesov2012optical, Pr3YSOsinglecenterSpec, kolesov2013mapping} and quantum memories
\cite{zhong2015optically, clausen2011quantum} 
to lasing materials \cite{moulton1982ti, maiman1960stimulated}. To exploit their potential in a 
scalable architechure, these substrates need to be enabled by a waveguiding platform. 
Previously demonstrated photonic structures in these materials based on femtosecond laser-writing
share common drawbacks like a low refractive index contrast between waveguide and substrate
\cite{calmano2010nd}. 
The corresponding increase of the minimum feature size of photonic elements by up to 
two orders of magnitude makes this technology unpractical for the field of cavity quantum 
electrodynamics (CQED) and nanoscale photonics in general. 

An alternative way to realize a waveguiding platform is the deposition of a
high refractive index thin film on top of these substrate crystals, which then provides on-chip 
access to these systems through evanescent light fields
\cite{vetsch2010optical, liebermeister2014tapered}
 coupling to embedded emitters. Moreover, this approach leaves the substrate crystal untouched 
 from processing
 and preserves the spectroscopic properties of embedded emitters \cite{marzban2015observation}.

In the visible range, titanium dioxide (TiO$_2$) features the highest refractive index 
out of a variety of transparent thin film materials, thus allowing for waveguiding on the 
majority of transparent substrates. Additionally, 
it has a wide transparency range covering the whole visible and near-infrared spectral regions.
Shaping the deposited TiO$_2$ film into photonic elements can then be conveniently done by reactive
ion etching (RIE) \cite{quidant2004addressing}. The deposited
TiO$_2$ films often tend to form
nano-crystallites which lead to significant scattering and, therefore, to substantial optical
losses \cite{abe2011nonlinear, alasaarela2013high}. The best performance is typically shown by
amorphous TiO$_2$ films featuring a refractive index between 2.3 and 2.4 
\cite{hayrinen2014low, alasaarela2013high, bradley2012submicrometer, choy2012integrated,
furuhashi2011development}.
In order to 
preserve the amorphous composition of the film, special precautions have to be taken.
\\
In the present work, we report on Chemical Beam Vapour Deposition (CBVD) of high optical 
quality Si:TiO$_2$ film whose amorphous state is preserved by doping with silicon
\cite{karasinski2011optical}. 
The amorphisation of deposited TiO$_2$ with Si doping was already 
reported \cite{wagner2016geometry}, and a more recent detailed study of
amorphisation of Nb:TiO$_2$ 
thin films by doping with Si is reported elsewhere \cite{sandu2016combinatorial}. 
Even though this doping leads 
to a slight decrease in the refractive index of the film, in agreement with results achieved
with other Chemical Vapour deposition techniques \cite{lee2000deposition}, the index 
is still high enough to support 
waveguiding on high-index optical crystals such as YAG and sapphire. On the other hand,
Si doping keeps the TiO$_2$ from forming nano-crystallites even at elevated temperatures 
up to 650$^{\circ}$C, resolving the inherent thermal instability of amorphous TiO$_2$ 
films to some extent \cite{martin1997microstructure}.

The optical performance of the film was assessed by 
fabricating
whispering gallery mode (WGM) cavities and measuring their Q-factor. The latter was in
excess of 10$^{5}$.
Evanescent coupling to fluorescent spots in the substrate material can be shown and also 
additional functionality can be added to the Si:TiO$_2$
film by doping it 
with fluorescent rare-earth ions, such as erbium. This shows the potential 
of Si:TiO$_2$ nanophotonic structures to also act as active elements within photonic devices. 
\\

\textbf{Deposition of Si:TiO$_2$ combinatorial film}\\
\begin{figure}[h]
\includegraphics[width=0.5\textwidth]{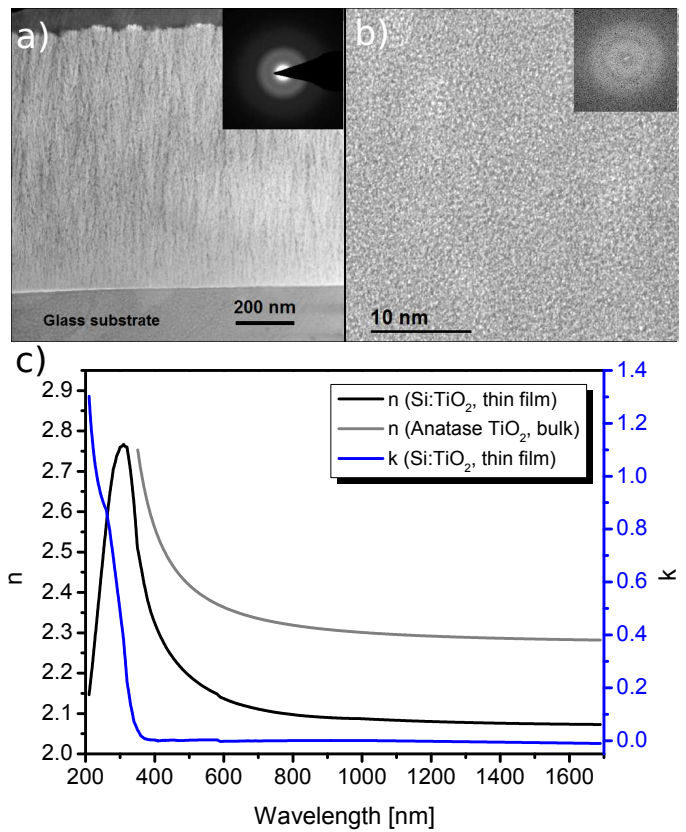}% Here is how to import EPS
\caption{Cross-sectional view by TEM of a Si:TiO$_2$ film deposited on glass substrate. 
a) HAADF-STEM image and SAED pattern. b) HRTEM image with inserted FFT. c) Refractive index
for an unpatterened Si:TiO$_2$ film on silica sample and for comparison the refractive 
index of bulk anatase TiO$_2$ \cite{devore1951refractive}. $n$ and $k$ denote the refractive index
and the extinction coefficient, respectively. 
\label{fig:img1}}
\end{figure}

Oxide thin films were deposited by CBVD, as described 
in detail elsewhere \cite{wagner2016geometry}. The CBVD process makes use of thermal
decomposition of organometallic 
precursors on a heated substrate in a high vacuum environment ($\leq10^{-3}$\,Pa).
These precursors impinge upon the substrate as molecular beams and do not undergo any
gas phase reaction. Deposition was conducted on epi-polished YAG crystals (CrysTec) and
quartz for functional characterization and on Si and glass wafers for material 
characterization. The liquid organometallic precursors used during the present
investigation were titanium tetraisopropoxide (Ti(OiPr)$_4$, CAS 546-68-9 evaporated from
a reservoir at 32$^{\circ}$C) and tetrabutoxysilane (Si(OnBu)$_4$, CAS 4766-57-8, 
evaporated from a 
reservoir at 65$^{\circ}$C). The Ti and Si precursor flows were homogeneously distributed 
on the
substrate (the flow ratio of precursors was estimated as Si/Ti=1.1). The substrate temperature 
during the deposition was kept at 500$^{\circ}$C. Under these conditions, the Si doping in the
film 
(defined as Si/(Si+Ti) ratio) was about 25\% (as measured by SEM-EDX) and the growth rate was
about $10 \pm 2$\,nm/min. Before starting the deposition, the chamber was pumped to a base 
pressure 
of 5$\times$10$^{-4}$\,Pa. A liquid nitrogen-cooled cryo-panel helped to maintain a pressure 
below 
2$\times$10$^{-3}$\,Pa during the deposition. 
The morphology and the chemical composition of the thin films were investigated by SEM-EDX 
using a Merlin SEM and in TEM cross-section using a Tecnai Osiris microscope. From 
cross-sectional TEM images, we can estimate the average thickness of deposited films
and their growth morphology. 
Typical images of characterized films are presented in Figure 1 a) and b) for films of thicknesses
$800 \pm 100$\,nm. A High Angle Annular Dark Field image (Figure \ref{fig:img1} a) shows a 
homogeneous
compact film with relatively low roughness. The inserted Selected Area Electron Diffraction
pattern together with the High Resolution TEM image (Figure \ref{fig:img1} b) confirm the
amorphous phase of the film.
Further characterization of the unpatterned Si:TiO$_2$ film of a thickness of 535\,nm
by ellipsometry yields the 
dispersion of the refractive index as shown in figure \ref{fig:img1} c). Within the 
transparency window of the film starting roughly at 400\,nm and extening all the way to the
infrared, the refractive index ranges between 2.3 and 2.1.
\\

\textbf{Structuring Si:TiO$_2$ films}
\\
\begin{figure}[h]
\includegraphics[width=0.5\textwidth]{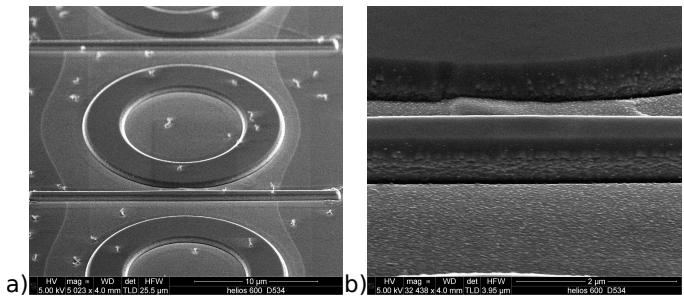}% Here is how to import EPS
\caption{a) SEM image of a Si:TiO$_2$ WGM resonator evanescently coupled to a straight optical 
waveguide. 
b) Close-up of the coupling region between ring resonator and access waveguide.
\label{fig:img2}}
\end{figure}

In order to assess the optical quality of the film, we fabricated
monolithic optical WGM resonators and tested the width of 
their resonances. The resonators were microrings evanescently coupled to a nearby optical 
waveguide through which excitation light was supplied.

The patterning was done by standard RIE with a Ni mask defined by e-beam lithography.
RIE of Si:TiO$_2$ was performed in an atmosphere of Ar/CF$_4$/O$_2$ with
the respective flow rates 4/16/3\,sccm and the RF power being 100\,W \cite{choi2013dry}. 
 The process pressure was 15\,mTorr and the total time required
to etch 
through 535\,nm of Si:TiO$_2$ was 24 minutes. After etching, the nickel mask was still present 
indicating that the etching selectivity was better than 1:11.
The residual nickel mask was removed by dissolving the metal 
in an aqueous 1M solution of nitric acid. In order to remove damage introduced into
Si:TiO$_2$ films by ion bombardment during plasma etching, the resulting structures were
annealed in air for 4 hours at 500$^{\circ}$C. However, annealing the film at 
temperatures above
800$^{\circ}$C leads to a visual change of the film, suggesting recrystallization.
Sample SEM images of the resulting structures 
are shown in Figure \ref{fig:img2}. The radius of the WGM resonator was 8\,$\mu$m while the
length of
the waveguide was 25\,$\mu$m. The distance between the resonator and the waveguide was around
400\,nm. 
\\

\textbf{Optical characterization of waveguides and resonators on various substrates}
\\
\begin{figure}[h]
\includegraphics[width=0.5\textwidth]{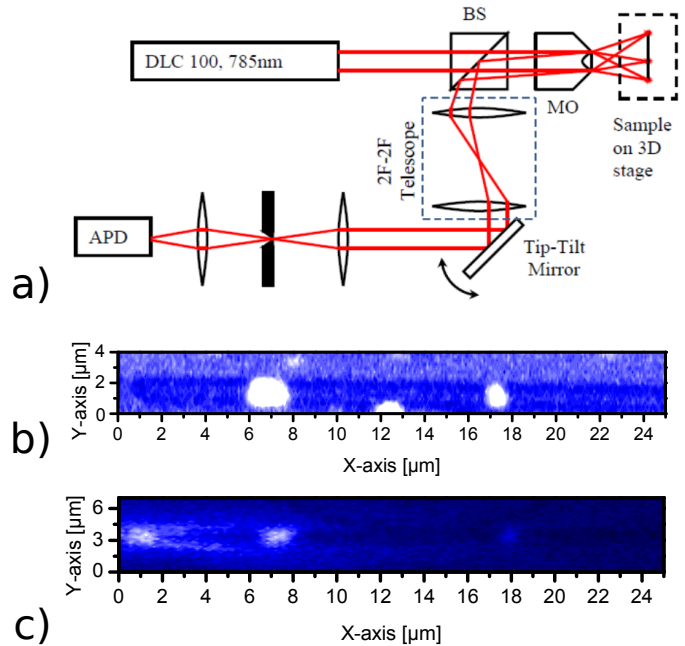}% Here is how to import EPS
\caption{a) Schematic diagram of the microscope. The sample is mounted on the 3D
nanopositioner and can be moved through the laser focus. Light emission is collected 
through the same objective lens and sent through a 2F-2F telescope and a 
tip-tilt mirror onto the pinhole selecting the observation point on the sample. b) Standard
confocal microscope scan with overlapping excitation and observation point. White spots
represent
fluorescence of Cr implanted spots in YAG, two of them underneath a fabricated Si:TiO$_2$
waveguide.
c) Tip-tilt mirror scan with
excitation point kept at a stationary position to couple light into the
fabricated Si:TiO$_2$ 
waveguide (left end). The two white spots at around 7\,$\mu$m and 18\,$\mu$m on the X-axis 
represent fluorescence of Cr implanted spots, evanescently excited by 
light traveling through the waveguide. 
\label{fig:img3}}
\end{figure}

In the following, waveguides and resonators were fabricated out of Si:TiO$_2$ thin films
on two different substrates, namely YAG and silica with a refractive index of 1.82 and 
1.45 at 790\,nm, respectively. 
Studies of the optical performance of waveguides and cavities were performed in a 
home-built
confocal microscope with an additional ability of scanning the detection point in the 
vicinity of the position of the excitation laser focus \cite{kolesov2009wave}.
Its schematic is shown in Figure \ref{fig:img3} a).
The excitation
laser was focused onto the sample with a high numerical aperture objective lens 
(Olympus 0.95$\times$50). \\
A YAG crystal (Y$_3$Al$_5$O$_{12}$) was doped with chromium by ion implantation through a 
perforated copper mask with an average hole diameter of $\approx$400\,nm. With an energy of
100\,keV and a dose of 10$^{12}$\,cm$^{-2}$, Cr ions end up in the YAG crystal in a depth of
$\approx 56 \pm 22$\,nm according to SRIM simulations \cite{ziegler2010srim}. A Si:TiO$_2$
film was 
subsequently deposited and shaped into waveguides in 
order to asses waveguiding and evanescent coupling of waveguided light to shallow
implanted
Cr-doped spots. Cr fluoresence of implanted spots could be detected close to 700\,nm
under
excitation with 600\,nm light as shown in the confocal scan in Figure \ref{fig:img3} b). 
Since the refractive index of YAG is smaller than the index of the film, waveguiding could be
observed. Incoupled light traveling
within the
waveguide was able to evanescently excite Cr implanted spots as shown in the
tilting mirror scan in
Figure \ref{fig:img3} c). Here, the excitation laser position was kept stationary 
coupling light into the left end of the waveguide. By scanning 
the point of detection with the tilting mirror, not only the excitation laser position 
yields signal, but also implanted Cr spots light up, which are embedded
below the waveguide extending to the right. 
This confirms the
evanescent coupling of light between fabricated waveguide and emitters in the substrate.
In combination with high quality resonators, this film-substrate interaction has the
potential to facilitate CQED experiments with rare-earth ion doped
crystals \cite{marzban2015observation}.

\begin{figure}[h]
\includegraphics[width=0.5\textwidth]{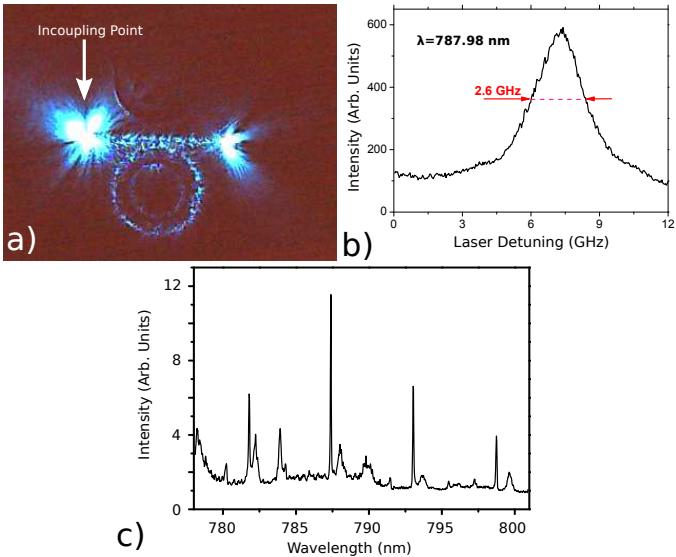}% Here is how to import EPS
\caption{a) Camera image of a structure excited on resonance. The left end of the 
waveguide was used as input and the cavity mode is visualized by the residual scattering
of the laser being in resonance. Tunable blue diode laser was used for visualization. b) 
The spectral shape of 
the cavity mode was obtained by sweeping a single-mode laser through the cavity resonance. 
 c) Typical mode spectrum of the studied resonator geometry between 780\,nm and 800\,nm obtained 
 with a broadband light source. Mode width measurement is limited by the resolution of the
 spectrometer for the fundamental mode.
\label{fig:img4}}
\end{figure}

\begin{figure}[h]
\includegraphics[width=0.5\textwidth]{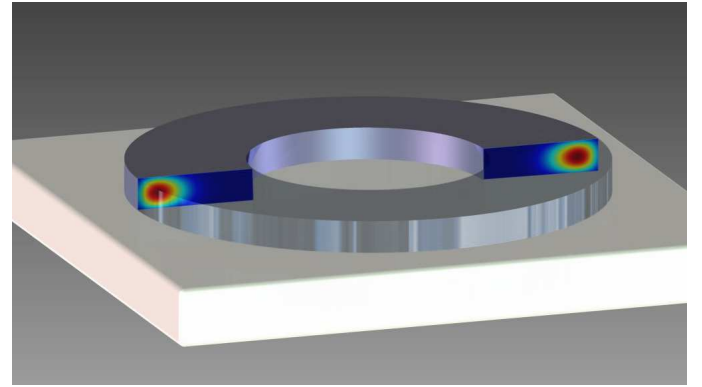}% Here is how to import EPS
\caption{Resonator sketch with calculated electric field profile of the fundamental 
790-nm mode.
\label{fig:img6}}
\end{figure}

For characterizing fabricated Si:TiO$_2$  resonators on glass substrates, the widths of the 
cavity 
resonances were studied with a single frequency tunable diode
laser (Toptica DL Pro). The laser could be tuned coarsely over the range of 770-800\,nm with 
a mode-hop-free fine tuning range of 30\,GHz. The spectral width of the laser was below 
1\,MHz. Laser output was inserted into one of the ends of the optical waveguide and its
frequency was swept while monitoring the scattered emission at the rim of the cavity. 
The rim lit up once the laser was in resonance with one of the cavity modes due to 
residual scattering on the imperfections of the structured film (see Figure \ref{fig:img4} a)). This scattering was 
detected as a function of laser frequency to estimate the spectral width of the mode.
The result of the spectral measurements is shown in Figure \ref{fig:img4} b).
The estimated mode width 
was 2.6\,GHz as the laser was finely swept around 787.89\,nm wavelength. This value 
corresponds to the quality factor of $Q\approx146000$. 
For this specific resonator geometry with an outer resonator radius
of $R=8\,\mu$m, a film thickness of 535\,nm and a rim width of 2\,$\mu$m, the mode spectrum
was measured with a broadband light source in a wavelength region
between 780\,nm and 800\,nm in order to extract
the free spectral range (FSR) of $\Delta \lambda_{\mbox{\tiny FSR}}=5.60$\,nm for the 
fundamental mode. 
Including the refractive index measurement, our corresponding finite element method based 
simulation can confirm the measured FSR for the fundamental mode
in this resonator geometry. The respective resonator sketch and the electric field profile
of the mode is shown in  
Figure \ref{fig:img6}.
With the FSR measurement around 790\,nm, the group index of 2.18 was estimated by
$n_g \approx \lambda^2/(2 \pi  R \cdot \Delta \lambda_{\mbox{\tiny FSR}})$. 
Based on the quality factor measurement and the group index estimation,
propagation losses $\alpha$
in fabricated waveguides of 5.1\,dB/cm were estimated by $\alpha = 2 \pi n_g /  Q \lambda$.

Table \ref{tab:table1} compares different TiO$_2$-based thin film waveguiding structures based 
on their
propagation loss. The doped Si:TiO$_2$ film reaches benchmark propagation losses in 
the visible in exchange for a doping dependent decrease of the refractive index. 

\setlength{\tabcolsep}{5mm}
\renewcommand{\arraystretch}{1.2}
\begin{table}[h!]
  \centering

  \caption{Comparison of propagation loss in dB/cm in TiO$_2$-based thin film waveguiding structures.}
  \label{tab:table1}
  \begin{tabular}{cccc}

  \hline
 Film &   Loss at  & Loss at & Reference \\
  details &   600-800\,nm &  1550\,nm&   \\
    \hline \hline
    
RF magnetron   &  9 & 4 & \cite{bradley2012submicrometer}\\
sputtering& & & \\
laser molecular &  57& - & \cite{abe2011nonlinear}\\
beam epitaxy & & &\\
atomic layer & -  & 2.4 & \cite{hayrinen2014low} \\
deposition& & &\\
RF magnetron  & 9.7 & - & \cite{furuhashi2011development}\\
sputtering& & &\\
CBVD  & 5.1 & - & this work
    
  \end{tabular}
\end{table}

\textbf{Doping of Si:TiO$_2$ resonators with erbium}\\

\begin{figure}[h]
\includegraphics[width=0.5\textwidth]{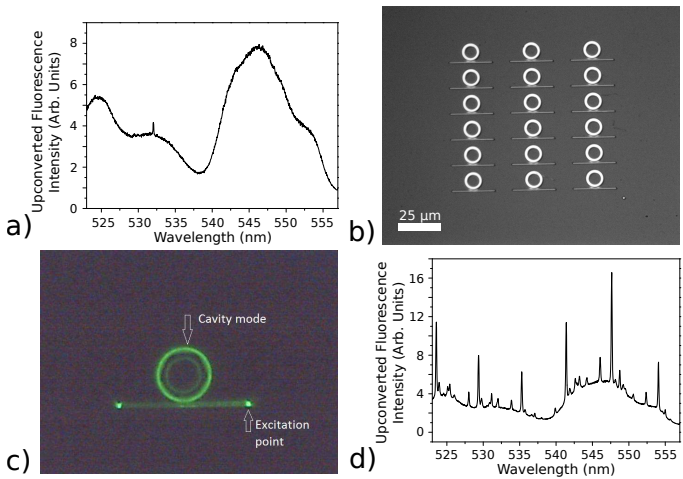}% Here is how to import EPS
\caption{a) Spectrum of upconverted fluorescence of erbium in Si:TiO$_2$ film. 
Excitation wavelength is 785\,nm. b) Microscope image of the WGM resonators manufactured 
out of Er$^{3+}$-implanted film. Scale bar is 25 $\mu$m long. c) Cavity mode visualized by upconverted fluorescence.
The
excitation laser is in resonance with one of the infrared cavity modes. d) Spectrum of 
erbium upconverted fluorescence collected at the end of the waveguide. 
The smooth background 
is due to the
fluorescence of the waveguide. The sharp peaks correspond to resonances of the cavity.
\label{fig:img5}}
\end{figure}

Another way of adding functionality to the thin film, specifically to the fabricated 
thin film resonators, is by optical activation with fluorescent 
species. We have chosen rare-earth (RE) doping due to robust optical properties of RE ions 
in most crystalline and glassy transparent media. RE doping can be performed by ion 
implantation
with very high yield of fluorescent species \cite{kornher2016production}. In addition,
optical materials doped with
erbium exhibit strong upconverted fluorescence in the visible once excited in the infrared.
This makes erbium doping conveniently detectable. The Si:TiO$_2$ film on glass was implanted
with 
erbium ions of 2\,MeV energy and with a dose of 10$^{14}$\,cm$^{-2}$. According to SRIM 
simulations, 
this leads to an erbium doping inside the Si:TiO$_2$ film in a depth of
$\approx 350 \pm 78$\,nm, 
corresponding to a maximum local density of $5 \cdot 10^{18}$\,cm$^{-3}$
\cite{ziegler2010srim}. Immediately after the
implantation,
the appearance of the film was changed from pink to grey, probably, due to 
implantation-induced
damage. At this point no upconverted fluorescence from Er$^{3+}$ ions was detected. 
Post-implantation
annealing in air at 500$^{\circ}$C for 4 hour heals out the 
implantation damage 
and, at the same time, activates erbium emission. After annealing, the film restored its 
original color. At the same time, strong green upconverted fluorescence of erbium ions 
could be detected under the excitation with red (650\,nm) and infrared 
(800\,nm)
laser light. The spectrum of the upconverted emission is shown in Figure \ref{fig:img5} a), in
good agreement with other works on erbium in glassy environment 
\cite{vetrone2002980, song2001three}.

The film was shaped to form whispering gallery mode cavities with 5\,$\mu$m radius coupled
to straight waveguides as described above (see Figure \ref{fig:img5} b). The 
upconversion resonances of 
Er$^{3+}$ in glassy environment are broad ($\approx$10\,nm), therefore, several infrared
cavity modes 
could lead to upconverted fluorescence. Once the infrared excitation laser is tuned in 
resonance with one of such modes, green fluorescence on the rim of the WGM cavity lights
up (see Figure \ref{fig:img5} c, the excitation laser is filtered out). 
The spectrum of the fluorescence collected at one of the waveguide 
ends 
shows its mode structure as indicated in Figure \ref{fig:img5} d).\\

\textbf{Conclusion}\\

We have demonstrated a method of depositing low loss high index Si:TiO$_2$ films
on different substrates such as glass, sapphire, and YAG. For most
substrates, the refractive index of the film is high enough to support waveguiding and 
also evanescently excite shallow fluorescent centers within the substrate material.
We have also shown a way of structuring the film to form on-chip photonic elements such
as waveguides and resonators. The low propagation loss of 5.1 db/cm results in a
high Q-factor of the resonators (1.5$\times$10$^{5}$ at 800\,nm) and underlines the potential for
CQED application in connection with rare-earth ion doped crystals for example.
Finally, due to the increased thermal stability of this film when compared to TiO$_2$, 
we could demonstrate further optical functionalization
of the film by doping with fluorescent rare-earth species (erbium). These results show how
the silicon doped titania film can be applied to on-chip photonics in various ways.
\\

\textbf{Acknowledgements}\\

The authors wish to thank the CIME-EPFL team for their TEM investigation facilities. 
The authors wish to acknowledge the FEDER (Fonds Europ\'{e}en de D\'{e}veloppement
Economique
et R\'{e}gional) for financing the Nanobium project through the Interreg IVA 
programme. The 
work was also financed by ERC SQUTEC, EU-SIQS SFB TR21, and DFG KO4999/1-1.

\bibliography{newbibfile}

\end{document}